\documentclass[aps,twocolumn,showpacs]{revtex4}
\usepackage{amsmath}
\usepackage{epsfig}

\begin{document}

\title{Hidden strange pentaquark in constituent quark models}

\newcommand*{\NJNU}{Department of Physics, Nanjing Normal University, Nanjing, Jiangsu 210097, China}\affiliation{\NJNU}

\author{Xuejie Liu}\affiliation{\NJNU}
\author{Hongxia Huang}\email{hxhuang@njnu.edu.cn}\affiliation{\NJNU}
\author{Jialun Ping}\email{jlping@njnu.edu.cn}\affiliation{\NJNU}

\begin{abstract}
In the framework of the chiral quark model, we investigate the hidden strange pentaquark system of the $N\phi$ state with the quantum numbers of IJ=$\frac{1}{2}$$\frac{3}{2}$. The results show that the $N\phi$ state can be bound through the interaction of the $\sigma$ meson exchange plus the effect of the channel coupling, which means that the effect of the channel coupling has an influence on the existence of this bound state.
\end{abstract}


\maketitle

\setcounter{totalnumber}{5}

\section{\label{sec:introduction}Introduction}
The quantum chromodynamics (QCD) is underlying theory of the strong interaction. However, it is difficult to study the structure of
the hadrons, and the hadron-hadron interaction directly because of the nonperturbative properties of QCD in the low energy region, although lattice QCD has made impressive progresses on nucleon-nucleon interactions and multiquark systems~\cite{qcdq,qcdw,qcde}.
The QCD-inspired quark models are still the main approach to study the hadron-hadron interactions and multiquark states.
The QCD does not forbid the existence of the exotic hadronic states other than $qqq$ baryons and $q\bar{q}$ mesons,
such as glueballs (without quark/antiquark), hybrids, multiquark states and hadron molecules. Searching of all these kinds of
states has a long history and various methods have been applied, such as the one-boson-exchange
(OBE) models~\cite{OBE}, the chiral perturbation theory~\cite{chpt}, the chiral quark models~\cite{quark_model}, and so on.
In 2015, the LHCb Collaboration announced the observation of the hidden charm pentaquarks $P_{c}(4380)$ and
$P_{c}(4450)$~\cite{lhcb,lhcba,lhcbb}. The discovery reignited the interests of theorists and experimentalists on the pentaquark
states~\cite{renew}. There are many theoretical analyses devoted to the hidden charm pentaquarks~\cite{p1,p2,p3,p4,p5,p6}.

Inspired by the hidden charm pentaquark states, people are also interested in the hidden strange pentaquark states,
which are composed of $qqqs\bar{s}$. In fact, as early as 1999, the evidences for a new baryon state with hidden strange and
a mass of $>1.8$ GeV was reported by SPHINX Collaboration in studying hyperon-kaon mass spectra in several proton diffractive reactions~\cite{sp}. In 2001, the $N\phi$ bound state was proposed by Gao, {\em et al.}~\cite{gao}, followed the idea of Brodsky~\cite{brodsky}. The calculation of the $N\phi$ system in Ref.~\cite{gao} shown that the QCD Van der Waals attractive force~\cite{qcd}, mediated by multigluon exchanges, can be strong enough to form the $N\phi$ bound state. It is also pointed out
that the subthreshold quasifree $\phi$ meson photoproduction inside a nuclear medium will enhance the probability for the formation
of the $N\phi$ bound state~\cite{gao}. Therefore, further theoretical and experimental investigations is needed to understand
dynamics of the pentaquarks with hidden strange.

To search for the $N\phi$ state experimentally, many groups carry out the study of the $\phi$ production in nuclei.
For example, the KEK-PS-E$325$ Collaboration~\cite{kek}, the LEPS Collaboration at Spring8/Osaka~\cite{leps},
the CLAS Collaboration at the Jefferson Lab~\cite{jafflab}, and J\"{u}lich group~\cite{juli}. Besides, the hidden strange molecular states are also supported by analyses of experimental data of the relevant photo-productions.
In Ref.~\cite{co}, it is pointed out that the $\phi$ meson should form bound states with all the nuclei considered by solving the Klein-Gordon equation with complex optical potentials. In Ref.~\cite{DCa,dcaa}, the $\phi$ meson properties in cold nuclear matter
are investigated by implementing resonant $N\phi$ interactions. Recently, the Belle Collaboration reported their searching for
the decay of $\Lambda_{c}^{+}$$\rightarrow$$\pi^{0}p\phi$ and~no significant signal was observed with an upper limit on the branching fraction of B($\Lambda_{c}^{+}$$\rightarrow$$\pi^{0}p\phi$)
 $<$$15.3$$\times$10$^{-5}$ at a $90\%$ confidence level~\cite{bell}.

According to the experimental information, in the vicinity of $2$ GeV, a $N\phi$ bound state is predicted in several models~\cite{xie,gao,huang,hf,hejun,latticee}. Xie and Guo studied the possible $\phi$p resonance in the $\Lambda_{c}^{+}$$\rightarrow$$\pi^{0}p\phi$ decay by considering a triangle singularity mechanism on the basis of the investigation of the hidden strange pentaquark~\cite{xie}. In Ref.~\cite{huang}, the $N\phi$ resonance state was obtained by channel-coupling in
the quark delocalization color screening model (QDCSM), and they also performed a Monte Carlo simulation of the bound state
production with an electron beam and a gold target, and found it was feasible to experimentally search for the $N\phi$ bound state
through the near threshold $\phi$ meson production from heavy nuclei. In Ref.~\cite{hf}, they showed that the $N\phi$ state
was a quasi-bound state by considering channel-coupling of $N\phi$ and $\Lambda K^*$ in the chiral quark model (ChQM).
In Ref.~\cite{hejun}, a bound state can be produced from the $N\phi$ interaction with spin parity $3/2^{-}$ after introduction
of a Van der Waals force between nucleon and $\phi$ meson. Moreover, the lattice QCD calculation also supports the existence
of such kind of bound state ~\cite{latticee}. Hence, it is worth it to make a systematical study of $N\phi$ bound states
from the difference way in both experiment and theory, which will deep our understanding about the hidden-strange pentaquarks.

It is obvious that there is no common flavor quark between $N$ and $\phi$, which is similar to the dibaryon state $N\Omega$.
In Ref.~\cite{Nomeig}, we studied the strange dibaryon $N\Omega$ in both the QDCSM and the ChQM, and the similar results
were obtained. This indicates the consistency of describing $N\Omega$ in these two quark models. It is interesting to investigate
if or not coincident results of the $N\phi$ state can be obtained in both models. As we mentioned above, we have studied the
$N\phi$ state in QDCSM. So in the present work, we will study the $N\phi$ state in ChQM and the effect of the channels
coupling will also be investigated. This work will be helpful to understand the interaction of the nucleon and a $\phi$ meson
and the coupled-channel effect.

This paper is organized as follows. First, the chiral quark model is introduced briefly. The results for the $N \phi$
state are shown in Sec.\uppercase\expandafter{\romannumeral2}, where some
discussion is presented as well. Finally, the summary is given in Sec.\uppercase\expandafter{\romannumeral3}

\section{THE CHIRAL QUARK MODEL}
The chiral quark model has acquired great achievement both in describing the hadron spectra, nucleon-nucleon interaction
and multiquark states~\cite{chiral}. In this model, the massive constituent quarks interact with each other through
Goldstone boson exchange in addition to one-gluon exchange. Besides, the color confinement and the scalar octet meson
exchange are also introduced. The Hamiltonian of ChQM for the present calculation takes the form:


\begin{widetext}
\begin{eqnarray}
H &=& \sum_{i=1}^{5} \left(m_i+\frac{p_i^2}{2m_i}\right)-T_CM+\sum_{j>i=1}^5\left(V_{ij}^C+V_{ij}^G+V_{ij}^\chi
+V_{ij}^\sigma\right),\\
V_{ij}^{C} &=& -a_{c}\boldsymbol{\lambda^c_{i}}\cdot\boldsymbol{\lambda^c_{j}}(r_{ij}^2+v_0),\\
V^{G}_{ij} &=& \frac{1}{4}\alpha_s \boldsymbol{\lambda}^{c}_i \cdot\boldsymbol{\lambda}^{c}_j
\left[\frac{1}{r_{ij}}-\frac{\pi}{2}\delta(\boldsymbol{r}_{ij})(\frac{1}{m^2_i}+\frac{1}{m^2_j}
+\frac{4\boldsymbol{\sigma}_i\cdot\boldsymbol{\sigma}_j}{3m_im_j})-\frac{3}{4m_im_jr^3_{ij}}
S_{ij}\right] \label{sala-vG} \\
V^{\chi}_{ij} & = & V_{\pi}( \boldsymbol{r}_{ij})\sum_{a=1}^3\lambda_{i}^{a}\cdot \lambda
_{j}^{a}+V_{K}(\boldsymbol{r}_{ij})\sum_{a=4}^7\lambda_{i}^{a}\cdot \lambda _{j}^{a}
+V_{\eta}(\boldsymbol{r}_{ij})\left[\left(\lambda _{i}^{8}\cdot
\lambda _{j}^{8}\right)\cos\theta_P-(\lambda _{i}^{0}\cdot
\lambda_{j}^{0}) \sin\theta_P\right] \label{sala-Vchi1} \\
V_{\chi}(\boldsymbol{r}_{ij}) & = & {\frac{g_{ch}^{2}}{{4\pi}}}{\frac{m_{\chi}^{2}}{{\
12m_{i}m_{j}}}}{\frac{\Lambda _{\chi}^{2}}{{\Lambda _{\chi}^{2}-m_{\chi}^{2}}}}
m_{\chi} \left\{(\boldsymbol{\sigma}_{i}\cdot\boldsymbol{\sigma}_{j})
\left[ Y(m_{\chi}\,r_{ij})-{\frac{\Lambda_{\chi}^{3}}{m_{\chi}^{3}}}
Y(\Lambda _{\chi}\,r_{ij})\right] \right.\nonumber \\
&& \left. +\left[H(m_{\chi}r_{ij})-\frac{\Lambda_{\chi}^3}{m_{\chi}^3}
H(\Lambda_{\chi} r_{ij})\right] S_{ij} \right\}, ~~~~~~\chi=\pi, K, \eta, \\
V^{\sigma}_{ij} & = & -{\frac{g_{ch}^{2}}{{4\pi }}}
{\frac{\Lambda _{\sigma}^{2}}{{\Lambda _{\sigma}^{2}-m_{\sigma}^{2}}}}%
m_{\sigma}\left[ Y(m_{\sigma}\,r_{ij})-{\frac{\Lambda _{\sigma}}{m_{\sigma}}}%
Y(\Lambda _{\sigma}\,r_{ij})\right] , \\
S_{ij}&=&\left\{ 3\frac{(\boldsymbol{\sigma}_i
\cdot\boldsymbol{r}_{ij}) (\boldsymbol{\sigma}_j\cdot
\boldsymbol{r}_{ij})}{r_{ij}^2}-\boldsymbol{\sigma}_i \cdot
\boldsymbol{\sigma}_j\right\},\\
H(x)&=&(1+3/x+3/x^{2})Y(x),~~~~~~
 Y(x) =e^{-x}/x. \label{sala-vchi2}
\end{eqnarray}
\end{widetext}
where $T_{CM}$ is the center of mass kinetic energy, and $\boldsymbol{\sigma}, \boldsymbol{\lambda}^c, \boldsymbol{\lambda}^a$
are the SU(2) Pauli, SU(3) color, SU(3) flavor Gell-Mann matrices, respectively. The subscripts $i$, $j$ denote the
quark index in the system. The $Y(x)$ and $H(x)$ are the standard Yukawa functions~\cite{chiral}, the $\Lambda_{\chi}$ is the
chiral symmety breaking scale, and the $\alpha_{s}$ is the effective scale-dependent running quark-gluon coupling
constant~\cite{oge}~,
\begin{equation}
\alpha_{s}(\mu)=\frac{\alpha_{0}}{\ln\left(\frac{\mu^2+\mu_{0}^2}{\Lambda_{0}^2}\right)}
\end{equation}
where~$\mu $~is the reduced mass of the interacting quarks pair.
$\frac{g_{ch}^2}{4\pi}$ is the chiral coupling constant for scalar and pseudoscalar chiral field coupling , determined from ~$\pi$~-nucleon-nucleon coupling constant through
\begin{equation}
\frac{g_{ch}^{2}}{4\pi}=\left(\frac{3}{5}\right)^{2} \frac{g_{\pi NN}^{2}}{4\pi} {\frac{m_{u,d}^{2}}{m_{N}^{2}}}
\end{equation}

To study the $u,d,s$ three flavor states, chiral SU(2) quark model has to be extended to chiral SU(3) quark
model~\cite{biaol,exbiaol}. There are two ways to introduce the scalar meson exchange. One is only $\sigma$ meson exchange
was used between any quark pair, another is introduce the full SU(3) scalar nonet meson exchange.
In this work, both ways are employed in the dynamical study the $N\phi$ interaction in the framework of resonating group
method (RGM). To check the model dependence of the results, we use three kinds of chiral quark models:
(1) ChQM1, the $\sigma$ meson exchange is used between any quark pair. (2) ChQM2, the $\sigma$ meson exchange is only valid
for the $u$, and/or the $d$ quark pair; (3) ChQM3, the full SU(3) scalar nonet-meson exchange is employed. These scalar nonet-meson exchange potentials~\cite{biaol} has the same form as the $\sigma$ meson exchange of the SU(2) ChQM, that is,
\begin{eqnarray}
V^{\sigma_{a}}_{ij} & = & V_{a_{0}}(
\boldsymbol{r}_{ij})\sum_{a=1}^3\lambda _{i}^{a}\cdot \lambda
_{j}^{a}+V_{\kappa}(\boldsymbol{r}_{ij})\sum_{a=4}^7\lambda
_{i}^{a}\cdot \lambda _{j}^{a} \nonumber \\
& & +V_{f_{0}}(\boldsymbol{r}_{ij})\lambda _{i}^{8}\cdot \lambda
_{j}^{8}+V_{\sigma}(\boldsymbol{r}_{ij})\lambda _{i}^{0}\cdot
\lambda _{j}^{0} \label{sala-su3} \\
V_{k}(\boldsymbol{r}_{ij}) & = & -{\frac{g_{ch}^{2}}{{4\pi }}}
{\frac{\Lambda _{k}^{2}m_{k}}{{\Lambda_{k}^{2}-m_{k}^{2}}}}%
\left[ Y(m_{k}\,r_{ij})-{\frac{\Lambda _{k}}{m_{k}}}%
Y(\Lambda _{k}\,r_{ij})\right] , \nonumber
\end{eqnarray}
with $k=a_{0}, \kappa, f_{0}$ and $\sigma$.

All the other symbols in the above expressions have their usual meanings, all the parameters needed in the present
calculation are taken from Ref.\cite{Nomeig} and listed in the Table~\ref{parameters}. The calculated baryon and meson masses
are presented in the Table~\ref{mass} with the experimental values.

\begin{table}[hb]
\caption{\label{parameters} The parameters of models:
$m_{\pi}=0.7$fm$^{-1}$, $m_{ k}=2.51$fm$^{-1}$,
$m_{\eta}=2.77$fm$^{-1}$, $m_{\sigma}=3.42$fm$^{-1}$,
$m_{a_{0}}=m_{\kappa}=m_{f_{0}}=4.97$fm$^{-1}$,
$\Lambda_{\pi}=4.2$fm$^{-1}$, $\Lambda_{K}=5.2$fm$^{-1}$,
$\Lambda_{\eta}=5.2$fm$^{-1}$, $\Lambda_{\sigma}=4.2$fm$^{-1}$,
$\Lambda_{a_{0}}=\Lambda_{\kappa}=\Lambda_{f_{0}}=5.2$fm$^{-1}$,
$g_{ch}^2/(4\pi)$=0.54, $\theta_p$=$-15^{0}$. }
\begin{tabular}{ccccc}
\hline\hline
              &                              & ChQM1        & ChQM2        &ChQM3\\   \hline
              &$b$(fm)                       &~~0.518       &~~0.518       &~~0.518    \\
              & $m_u$(MeV)                   &~~313         &~~313         &~~313      \\
              & $m_d$(MeV)                   &~~313         &~~313         &~~313      \\
              & $m_s$(MeV)                   &~~573         &~~536         &~~573      \\
              &$a_c$(MeV)                    &~~48.59       &~~48.59       &~~48.59    \\
              &$v_{0}$(MeV)                  &-1.2145       &-1.2145       &-0.961     \\

              &$\alpha_0$                    &~~0.510       &~~0.510      &~~0.583     \\
              &$\Lambda_0($fm$^{-1})$        &~~1.525       &~~1.525      &~~1.616     \\
              &$\mu_0$(MeV)                  &445.808       &445.808      &422.430     \\

\hline\hline
\end{tabular}
\end{table}

\begin{table}[ht]
\caption{\label{mass}The masses of ground-state baryons and mesons (in MeV).}
\begin{tabular}{lccccccc}
\hline \hline
               & ~~$N$~~            & ~~$\Lambda$~~ & ~~$\Sigma$~~
               & ~~$\Sigma^*$~~     & ~~$\phi$~~    & ~~$K$~~   & ~~$K^*$~~  \\ \hline

ChQM1      & 939          & 1124    & 1239
           & 1360         & 1056    & 695  & 817              \\
ChQM2      & 939          & 1137    & 1245
           & 1376         & 1016    & 686   & 817          \\
ChQM3      & 939          & 1123    & 1267
           & 1344      &1054  &667 &840       \\ \hline
Expt       & 939   &1116 &1193 &1385   &1019 &498 &892   \\  \hline\hline
\end{tabular}
\end{table}

\section{The results and discussions}

Here, we investigate the properties of the $N\phi$ state with the quantum numbers of $IJ=\frac{1}{2}\frac{3}{2}$ within
the chiral quark models mentioned above. The main goal of this work is to investigate whether the $N\phi$ state is bound or
not in ChQM. A dynamic calculation based on the resonating group method (RGM), a well established method for studying the
bound or scattering five-quark states, was done here. Expanding the relative motion wave function between two cluster by
employing well-defined basis wave functions, such as Gaussian functions, the integro-differential equation of RGM
transforms into algebraic equation-a generalized eigen-equation. Then one can solve the generalized eigen-equation
for a bound-state problem and obtain the corresponding binding energy. The details of solving the RGM equation
can be found in Ref.~\cite{KW,MK,MOKA}.

In our calculation, we consider the effect of the channel-coupling including both the color-singlet channels and the
hidden-color channels. The color-singlet means the color symmetries of $qqq$ cluster and $q\bar{q}$ cluster are all
[111], and the hidden-color indicates the color symmetries of them are all [21]. The colorful $qqq$ cluster and
$q\bar{q}$ cluster are listed in Table~\ref{colorful} and Table~\ref{colorful1}, respectively. The color symmetry $[c]$,
spin symmetry $[\sigma]$, flavor symmetry $[f]$, isospin $I$, and strangeness $S$ of the $qqq$ and $q\bar{q}$ clusters
are listed in these two tables. Moreover, the labels of all $16$ channels of the $N\phi$ system are listed in
Table ~\ref{channels}. The baryon-meson separation is taken to less than $6$ fm for the calculation of this system.
\begin{center}
\begin{table}[ht]
\caption{\label{colorful}The symmetries of colorful $qqq-$cluster.}
\begin{tabular}{lccccccccccccc}
\hline \hline
        & ~~$\Delta^{\prime}$~~  & ~~$\Sigma^{*\prime}$~~  & ~~$\Xi^{*\prime}$~~
         & ~~$N^{\prime\prime}$~~   & ~~$\Lambda^{\prime\prime}$~~  & ~~$\Sigma^{\prime\prime}$~~   & ~~$\Lambda^{\prime}_{s}$~~    \\ \hline
$[c]$   & $[21]$      & $[21]$       & $[21]$     & $[21]$  & $[21]$    &  $[21]$
     & $[21]$         \\
$[\sigma]$     & $[21]$       & $[21]$   & $[21]$     & $[3]$  & $[3]$
        & $[3]$   & $[21]$         \\
$[f]$      & $[3]$    & $[3]$       & $[3]$     & $[21]$& $[21]$
  & $[21]$  & $[111]$         \\
$I$  & $\frac{3}{2}$        & 1    & $\frac{1}{2}$     & $\frac{1}{2}$
        & 0    &  1    &  0       \\
$S$  & 0  & 1 & 2  & 0  & 1  & 1 & 1 \\  \hline\hline
\end{tabular}
\end{table}
\end{center}

\begin{center}
\begin{table}[ht]
\caption{\label{colorful1}The symmetries of colorful $q\bar{q}$ cluster.}
\begin{tabular}{lcccc}
\hline \hline
        & ~~$\eta^{\prime}$~~  & ~~$\phi^{\prime}$~~  & ~~$K^{*\prime}$~~
            & ~~$K^{*\prime}_{s}$~~    \\ \hline
$[c]$   & $[21]$      & $[21]$       & $[21]$     & $[21]$           \\
$[\sigma]$     & $[11]$       & $[2]$   & $[11]$     & $[2]$       \\
$[f]$      & $[21]$    & $[21]$       & $[21]$     & $[21]$         \\
$I$  & $0$        & 0    & $\frac{1}{2}$     & $\frac{1}{2}$    \\
$S$  & 2  & 2 & 1  & 1   \\  \hline\hline
\end{tabular}
\end{table}
\end{center}

\begin{table}[ht]
\caption{\label{kkk}Channels of the $N\phi$ system.}
\begin{tabular}{cccccccc}
 \hline \hline
  1 & 2 & 3 & 4 & 5 & 6  & 7  & 8 \\
 $N\phi$ &$\Lambda K^{*}$ &$\Sigma K^{*}$ &$\Sigma^{*}K$  &$\Sigma^{*}K^{*}$  &$\Delta^{\prime}\phi^{\prime}$ &$\Sigma^{*\prime}K^{*\prime}$  &$\Xi^{*\prime}K^{*\prime}$  \\ \hline
  9  & 10  & 11  & 12  & ~~13  & ~~14  & ~~15  & ~~16 \\
 $N^{"}\eta^{\prime}$  &$N^{"}\phi^{\prime}$  &$\Lambda^{"}K^{\prime}$  &$\Lambda^{"}K^{*\prime}$ &$\Sigma^{"}K^{\prime}$
  &$\Sigma^{"}K^{*\prime}$  &$\Sigma^{*\prime}K^{*\prime}$ &$\Lambda_{s}^{\prime}K^{*\prime}$                                  \\
 \hline\hline
\end{tabular}
\label{channels}
\end{table}

\begin{figure*}
\epsfxsize=6.5in \epsfbox{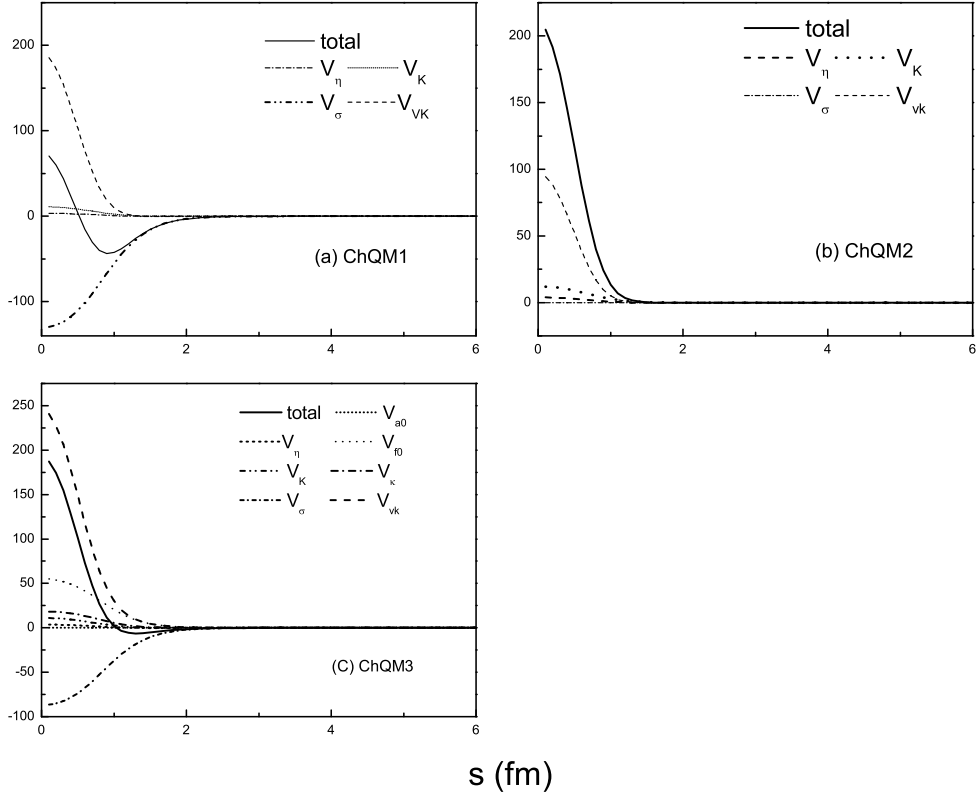} \vspace{-0.3in}                 \caption{The contributions to the effective potential from various terms of interactions.}
\end{figure*}

Since an attractive potential is necessary for forming bound state or resonance, the effective
potentials between $N$ and $\phi$ in three quark models are calculated and shown in Fig. 1. The effective potential
between two colorless clusters is defined as, $V(s)=E(s)-E(\infty)$, where $E(s)$ is the diagonal matrix element
of the Hamiltonian of the system in the generating coordinate. One sees that the potentials of $N\phi$ state in
both ChQM1 and ChQM3 are attractive, while the one in ChQM2 is repulsive. Besides, the attraction in ChQM1 is much
larger than the one in ChQM3.

To investigate the contribution of each interaction term to the total effective potentials between $N$ and $\phi$,
we also calculate the contribution of the kinetic energy ($V_{vk}$), the one-boson exchange ($V_{\pi}$, $V_{k}$ and
$V_{\eta}$), and the scalar octet-meson exchange($V_{a_0}$, $V_{f_{0}}$, $V_{\kappa}$, $V_{\sigma}$). Here, the
interactions of the one-gluon exchange ($V_{oge}$), the confinement ($V_{con}$) term, the one-pion exchange and
the $a_{0}$-meson exchange do not work between $N$ and $\phi$, because there have no common flavor quark between
these two clusters. The contributions of other terms to the effective potential are show in Figs.1(a)-1(c).
In ChQM1, the potential of the $\eta$, $K$ and the kinetic energy($V_{vk}$) are all repulsive. The $\sigma$ meson exchange
interaction is attractive. The total potential is attractive. This means it is possible to form $N\phi$ bound state in ChQM1.
In ChQM2, where $\sigma$ meson is restricted to exchange between u and the d quark pair only, this indicates that the
$\sigma$ meson exchange potential would have no contribution to the $N\phi$ state. It can be found that from the
Fig.$1$(b) that the total potential is repulsive. In ChQM3, as can be seen in Fig.1(c), although $\sigma$ meson
exchange potential yields a great deal of attraction, the potential of the $\eta$, $K$, $f_{0}$, $\kappa$ and
the kinetic energy are all repulsive. Their repulsive are sufficient to counteract the attraction of the $\sigma$
meson exchange interaction which causes very weak attraction of the total potential of the $N\phi$ state.
All these properties are consistent with our former study of $N\Omega$ state~\cite{Nomeig}, where no common
flavor quark between $N$ and $\Omega$.

\begin{table}[ht]
\caption{The binding energies B with channel-coupling.}
\begin{tabular}{lccc}
\hline \hline
  & ~$B_{sc}~(MeV)$~ & ~$B_{5cc}~(MeV)$~ & ~$B_{16cc}~(MeV)$~    \\ \hline

 ChQM1 & ub & -5.7 & -12.3  \\
 ChQM2 & ub & ub   & ub  \\
 ChQM3 & ub & ub   & ub  \\
 \hline
\end{tabular}
\label{bound1}
\end{table}

In order to see whether or not there is any bound state, we continue to do a dynamic calculation. The binding energies of the $N\phi$ state are listed in Table ~\ref{bound1}, where the $B_{sc}$ means the binding energy of the single-channel calculation of the $N\phi$ state; the $B_{5cc}$ express that the binding energy is produced by the coupling of five color-singlet channels; $B_{16cc}$ stands for the binding energy of all the 16 channels coupling. From Table~\ref{bound1} one can see that the single $N\phi$ is unbound (labelled as ``ub" in Table ~\ref{bound1} ) in all three quark models. This means that the attraction between $N$ and $\phi$ in ChQM1 is not large enough to from bound state; also the attraction between $N$ and $\phi$ in ChQM3 is too weak to make the $N\phi$ bound. Besides, it is also reasonable for the unbound $N\phi$ state in ChQM2 because of the repulsive interaction between $N$ and $\phi$.

Then we take into account the effects of the channel-coupling. From Table~\ref{bound1}, we can see that in ChQM1, the $N\phi$ state can be bound by the coupling of the color-singlet channels. Moreover, the binding energy can be larger by the coupling of the hidden-color channels. This means that the channel-coupling has an impact on the $N\phi$ state. This results are consistent with our former work of $N\phi$ state in the quark delocalization and color screening model QDCSM~\cite{huang}, in which the $N\phi$ state was bound by channel-coupling. However, the effect of the channel-coupling still not enough to generate the binding energy of the N$\phi$ state in both ChQM2 and ChQM3.





\

\section{Summary}

In this paper, we search for the $N\phi$ bound state within the chiral quark models: ChQM1, ChQM2 and ChQM3, which refer to the difference kinds of the $\sigma$ meson exchange. The calculated results show: (1) the potentials of the $N\phi$ state in both ChQM1 and ChQM3
are attractive, while the one in ChQM2 is repulsive. The attraction in ChQM1 is much larger than the one in ChQM3. (2) Although the single $N\phi$ is unbound in ChQM1, a bound $N\phi$ state can be obtained by the effect of channel-coupling. This means that the channel coupling effect in this model has an influence on the existence of the bound state. (3) The $N\phi$ is obviously unbound in ChQM2 because of the repulsive interaction between $N$ and $\phi$. (4) In ChQM3, the attraction between $N$ and $\phi$ is too weak to make the $N\phi$ bound.

The hidden charm pentaquark candidates have been reported by LHCb, we also expect the existence of the hidden strange pentaquark $N\phi$ state. The theoretical study of the pentaquark systems can help us to understand the nature of the multiquark states. The experimental search for the pentaquak systems will not only test the quark model, but also help us to understand the strong interaction. In the coming years, more and more novel phenomena
 are expected with experimental progress, expecially from LHCb and Belle$\Pi$. Clearly, it deserves the efforts from both theories and experiments.

\acknowledgments{}
This work is supported partly by the National Science Foundation
of China under Contract Nos. 11675080, 11775118 and 11535005, the Natural Science Foundation of
the Jiangsu Higher Education Institutions of China (Grant No. 16KJB140006).

\end{document}